\documentclass[conference,a4paper]{IEEEtran}

  	\usepackage[pdftex]{graphicx}
        \usepackage[tight,footnotesize]{subfigure}
  	\graphicspath{{../pdf/}{../jpeg/}}
	\DeclareGraphicsExtensions{.pdf,.jpeg,.png}
	\usepackage[cmex10]{amsmath}
	\usepackage{mathabx}
	\usepackage{algorithmic}
	\usepackage{array}
        \usepackage{tabularx}
	\usepackage{mdwmath}
	\usepackage{mdwtab}
	\usepackage{eqparbox}
	\usepackage{url}
        \usepackage{romannum}
        \usepackage{xcolor}
        \usepackage{balance}
        \usepackage{cite}
        \usepackage{geometry}
\begin{document}

\title{Exploring RIS Coverage Enhancement in Factories: From Ray-Based Modeling to Use-Case Analysis}


\author{\IEEEauthorblockN{
Gurjot Singh Bhatia\IEEEauthorrefmark{1}\IEEEauthorrefmark{2},   
Yoann Corre\IEEEauthorrefmark{1}, 
Thierry Tenoux\IEEEauthorrefmark{1}, 
M. Di Renzo\IEEEauthorrefmark{2}}                                     
\IEEEauthorblockA{\IEEEauthorrefmark{1} SIRADEL, Saint-Gregoire, France \\
\IEEEauthorrefmark{2} Universit\'e Paris-Saclay, CNRS, CentraleSup\'elec, Laboratoire des Signaux et Syst\`emes, Gif-sur-Yvette, France\\
gsbhatia@siradel.com}
}

\maketitle

\begin{abstract}
Reconfigurable Intelligent Surfaces (RISs) have risen to the forefront of wireless communications research due to their proactive ability to alter the wireless environment intelligently, promising improved wireless network capacity and coverage. Thus, RISs are a pivotal technology in evolving next-generation communication networks. This paper demonstrates a system-level modeling approach for RIS. The RIS model, integrated with the Volcano ray-tracing (RT) tool, is used to analyze the far-field (FF) RIS channel properties in a typical factory environment and explore coverage enhancement at sub-6 GHz and mmWave frequencies. The results obtained in non-line-of-sight (NLoS) scenarios confirm that RIS application is relevant for 5G industrial networks.
\end{abstract}

\vskip0.5\baselineskip

\IEEEpeerreviewmaketitle


\section{Introduction}

Reconfigurable Intelligent Surfaces (RISs) have arisen as a promising technology for enhancing wireless communication networks. RISs are usually nearly passive surfaces composed of a large number of sub-wavelength unit cells that can dynamically modify the phase and amplitude response of incident signals. This capability allows RISs to manipulate the wireless propagation environment, enabling various applications and performance improvements in wireless networks \cite{ETSI}.

RISs offer several advantages over traditional wireless communication systems. Firstly, RISs can shape the wireless channel by controlling the phase and amplitude of the reflected signals to enhance the wireless network's capacity and spectral efficiency, improve signal quality, and mitigate interference \cite{ETSI}. Secondly, RISs, unlike conventional relaying systems, do not require power amplifiers and can shape the incoming signal using phase shifts, making them energy-efficient and environmentally friendly \cite{ETSI, RIS1}. They seamlessly integrate with existing wireless networks, supporting full-duplex and full-band transmission \cite{ETSI, RIS1}.

RIS applications in wireless networks are diverse. They bolster cellular coverage, reduce latency in heterogeneous networks, and multi-access edge computing (MEC) \cite{RIS2}. RISs can aid in managing massive connectivity and interference in device-to-device communication. Additionally, they complement emerging technologies like unmanned systems, non-orthogonal multiple access (NOMA), and machine learning-based physical layers \cite{RIS2}, ushering in the next era of networks.



RIS modeling is critical to fully harness the benefits of RISs and explore the applications of RIS-assisted wireless communications. To actively address some of the challenges of RIS modeling, such as path-loss characterization, realistic environment modeling, and frequency dependence, this paper presents a system-level modeling technique for RIS, which is integrated with the Volcano ray-tracing (RT) \cite{Siradel} tool. The results from the RT tool are validated against measurement results in \cite{tang2022path}. Then, this ray-based system-level RIS propagation model is used to investigate and quantify coverage extension in a highly cluttered factory area at 3.7 GHz and 27 GHz. To the best of our knowledge, there has not been any such attempt for industrial environments.


This paper is structured as follows. Section II briefly explains the RIS modeling approach and validation. Section III demonstrates the RIS propagation model application. Section IV presents the results and analysis. Finally, Section V gives the conclusion and future perspective of this work.


\section{RIS-based Channel Modeling}

Different methods for modeling RIS surfaces have been suggested in the existing literature. A comprehensive model based on the Huygens-Fresnel principle and Green's theorem is available in \cite{RIS2, RIS4}. Some RIS path-loss models from the perspective of link-budget can be found in \cite{tang2020wireless, tang2022path, ellingson2021path}.

\vskip-0.3\baselineskip
\begin{figure}[ht!]
\centering
\includegraphics[width=2.2in]{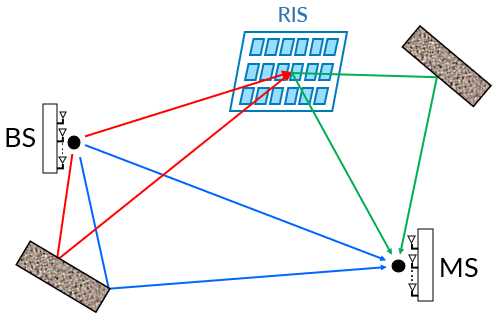}
\caption{Multi-path propagation scenario with RIS.}
\label{1.1}
\end{figure}

This paper assumes that the RIS is composed of discretized sub-wavelength sized elements called unit cells (UCs). Considering firstly that the network is composed of one base station (BS), one RIS, and one mobile station (MS) as shown in Fig. \ref{1.1}, the main goal is to compute the channel matrix $H_{w\_ris}$ between the BS and the MS, including the RIS. The channel matrix for a given path can be written as \cite{najafi2020physics}:

\vskip-0.5\baselineskip
\begin{equation}\label{eq1}
    H_{w\_ris} = H_{wo\_ris} + H_b \Omega H_a 
\end{equation}

where $H_{wo\_ris} (N\times M)$ is the channel matrix when no RIS is present (i.e., the direct channel). $N$ and $M$ are the number of antenna elements (AE) at the MS and the BS, respectively. Let the total number of RIS UCs be $K$. Then, $H_a (K\times M)$ is the channel matrix of the link $BS \rightarrow RIS$. $H_b (N \times K)$ is the channel matrix of the link $RIS \rightarrow MS$. 

It is assumed that the electromagnetic (EM) coupling between the UCs is negligible. Hence, $\Omega$ is a complex diagonal matrix describing the scattering properties of RIS UCs via their complex Radar Cross Section (RCS) \cite{najafi2020physics}: $g_{uc}$, and the configurable reflection coefficient $\Gamma_{uc}  = \tau e^{j\beta}$. Thus, the $\Omega$ matrix for the $n^{th}$ UC is given by:

\vskip-0.5\baselineskip
\begin{equation}\label{eq2}
    \Omega (n,n) = g_{uc} (n) \cdot \tau(n) e^{j\beta(n)} 
\end{equation}

\subsection{RIS UC scattering model}

There is a consensus to represent RIS as a discrete grid of UCs and sum coherently their contributions \cite{tang2020wireless, tang2022path, ellingson2021path}, but not necessarily about the form of UC scattering function $g_{uc}$. The $g_{uc}$ expressions presented in Table \ref{tab1} are not given in such form in \cite{tang2020wireless, tang2022path, ellingson2021path}, but were reformulated (based on the radar equation \cite{richards2014fundamentals}) from the given empirical path-loss expressions. Let $\theta_i$ denote the angle between the incident path and RIS normal, $\theta_s$ denote the angle between the scattered path and RIS normal, $d_x$ and $d_y$ denote the UC widths along the horizontal and vertical axes, respectively, and $G$ denote the maximum gain of the UC.

\vskip-0.4\baselineskip
\begin{table}[ht!]
 \centering
 \caption{UC scattering functions derived from state-of-the-art}
 \label{tab1}
 \begin{tabular}{|>{\centering\arraybackslash\vspace{0.2ex}}m{1.8cm}<{\vspace{0.5ex}}|>{\centering\arraybackslash\vspace{0.2ex}}m{5.5cm}<{\vspace{0.5ex}}|} 
  \hline
  Tang \textit{et al.}, eq. (1), (3) in \cite{tang2020wireless} & \vskip-0.3\baselineskip $g_{uc} = \left[d_x \ d_y \ G \ F(\theta_i) \ F(\theta_s)\right]^\frac{1}{2}$ \newline \newline $F(\theta)=(\cos \theta)^\alpha$ \& $G = 2(\alpha + 1); \alpha \geq 0$  \\ 
  \hline
  Tang \textit{et al.} \cite{tang2022path} & \vskip-0.2\baselineskip $g_{uc} = \frac{\left(4\pi \ \cos\theta_i \ \cos\theta_s\right)^\frac{1}{2} \ d_x \ d_y}{\lambda}$ \newline \newline $\alpha = 1 \Rightarrow F(\theta)=(\cos \theta)$ \& $G = \frac{4\pi \ d_x \ d_y}{\lambda^2}$  \\ 
  \hline
  Ellingson \cite{ellingson2021path} & \vskip-0.3\baselineskip $g_{uc} = G \ \lambda \left[ \frac{F(\theta_i) \ F(\theta_s)}{4\pi}\right]^\frac{1}{2}$ \newline \newline $F(\theta)=(\cos \theta)^\alpha$ \& $G = 2(\alpha + 1); \alpha \geq 0$ \\  
  \hline
 \end{tabular}
\end{table}

The maximum power radiated by the RIS in \cite{tang2022path} is the same as a metallic plate of the same size. The UC scattering model derived from \cite{tang2022path} shows good agreement with experimental results; that is why it has been selected for the present study. If the same angles and distances are considered for all UCs, then the model applies to only FF scenarios. We adopt this assumption as our foundational premise for this paper and concentrate on RIS far-field (FF) beamforming (BF). 

The RIS configuration is optimized according to the strongest path geometry. The reflection coefficients of all the UCs share the same amplitude value ($\tau(n) = A$). The design method for programmable phase shifts $\beta(n)$ is given by (9) in \cite{tang2020wireless} and is also used in \cite{tang2022path}. The parameters $\theta_i$ and $\theta_s$ are extracted from the RT tool. The computation of the matrices $H_a, H_b$, and $H_{wo\_ris}$ is carried out in the form of single-input single-output (SISO) links followed by multiple-input multiple-output (MIMO) extrapolation using the functions already available in the Volcano RT tool. 

The above-mentioned approach can be extended to calculate the $H$ matrices for several BSs and MSs, for a given RIS. The MIMO calculation is not further explained in this paper. A simplified setup with single-element antennas at both the BS and MS ends is considered instead. 

\subsection{RIS propagation model validation}

In order to validate the RIS propagation model, the measurement system A illustrated in \cite{tang2022path} is mimicked in the RT tool at 27 GHz to validate the direct line-of-sight (LoS) BS-RIS-MS link calculation. The measurement system A is used to measure the power distribution of the signal reflected from the RIS as a function of the angle of reception. The transmitting horn antenna and the RIS are fixed on the rotation platform in an anechoic chamber, and the incident signal is perpendicular to the RIS (normal incidence). When the platform is rotated, the receiver, which is constituted by a receiving horn antenna and an RF signal analyzer, can measure the received power at different angles of reception.

 
The received signal powers in primary, secondary, and tertiary lobes, emulated using $g_{uc}$ derived from \cite{tang2022path}, are -57.6 dBm, -71.2 dBm, and -75.7 dBm, respectively. These powers show good agreement with the theoretical and experimental results shown in Fig. 9 in \cite{tang2022path}. Then, we checked the LoS direct-path received signal powers after integration of the RIS model into the RT tool; the error is less than 0.8 dB. These results enhance the confidence in the accuracy of the RIS contributions simulated by the RT tool. The next step is to apply the RT tool to a more complex environment. It can be used to estimate and analyze channel parameters such as path-loss, time delay, optical visibility, delay spreads, etc. 


\section{RIS propagation model application}



Unlike propagation in typical home or office settings, propagation in factory scenarios is more site-specific. Strong obstructions in the LoS and fading fluctuations may reduce the reliability of the system and the signal. For instance, the huge bodies of metallic equipment and storage racks can act as significant obstacles. In such a scenario, RISs can be used to extend the coverage to the shadowed areas.  Besides coverage enhancements, RISs can enhance capacity, positioning, sensing, security, sustainability, and support several other use cases \cite{ETSI, RIS1, RIS2} for future smart factory scenarios.

\begin{figure}[ht!]
\centering
  \subfigure[BS1 coverage map at 3.7 GHz and the target area A1.]{%
    \includegraphics[width=2.8in]{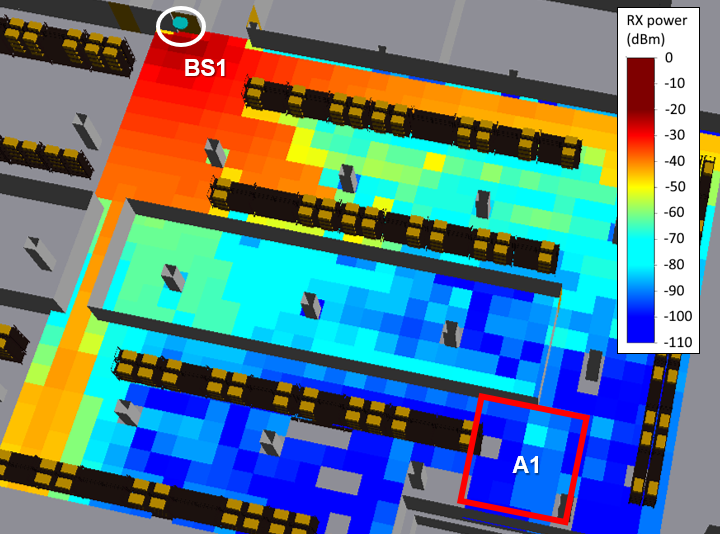}%
    \label{1.3.1}%
  } \\
  \subfigure[BS2 coverage map at 27 GHz and the target area A2.]{%
    \includegraphics[width=2.8in]{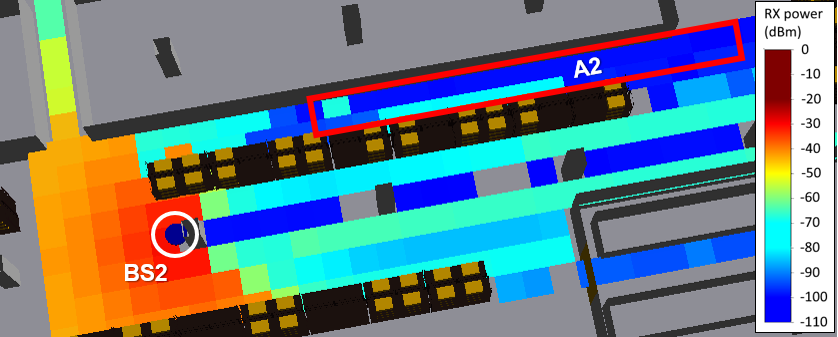}%
    \label{1.3.2}%
  }
  \caption{Coverage maps at 3.7 GHz and 27 GHz and the target areas.}
  \label{1.3}
\end{figure}

The propagation scenario for this study is the warehouse of a washing industry (see our previous work \cite{EuCNC1}), which is used in the Volcano RT tool with the RIS model to explore RIS-aided coverage enhancement at 3.7 GHz and 27 GHz. The height of the walls is 5 m, and that of storage racks is 4.4 m. The BSs are deployed at a height of 4 m, and the receiving MSs at a height of 1.5 m. In this analysis, the BS and MS antennas are vertically polarized single-element isotropic antennas; the transmitted power ($P_t$) at the BS end is 30 dBm. Fig. \ref{1.3.1} and Fig. \ref{1.3.2} show the BS1 coverage map at 3.7 GHz and the BS2 coverage map at 27 GHz, respectively. For this study, we selected the respective target areas A1 and A2 (marked with a red boundary) that show poor coverage levels due to strong obstructions.

\begin{figure}[ht!]
\centering
\includegraphics[width=3.05in]{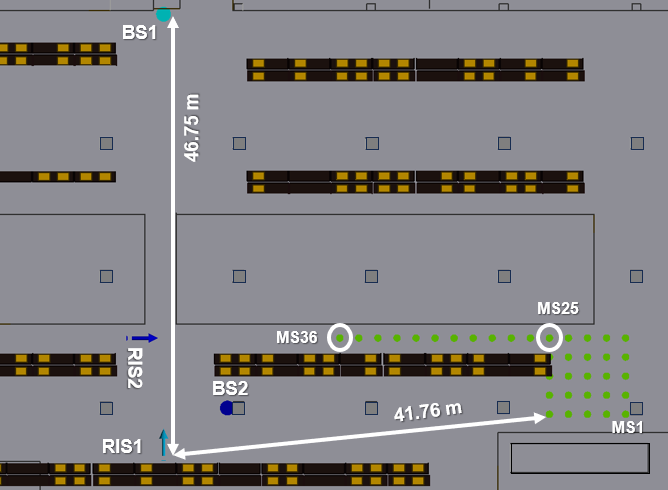}
\caption{BSs, MSs, and RIS deployment in the factory scenario.}
\label{1.5}
\end{figure}

The objective is to investigate how the coverage improves when using an RIS in LoS from the BS. Note that areas A1 and A2, as well as the RIS positions, have been chosen such that RIS FF conditions are met. RIS1 and RIS2 are deployed, as shown in Fig. \ref{1.5}. RIS1 is used at 3.7 GHz with BS1 to target A1, while RIS2 is used at 27 GHz with BS2 to target A2. Fig. \ref{1.5} also shows some distances for reference. The distance from BS2 to RIS2 is 13.19 m, and from RIS2 to MS36 is 22.52 m. RIS1 and RIS2 both are varied with respect to the number of UCs, but the RIS UC size: $d_x = d_y = \lambda / 2$ and $\lvert \Gamma_{uc} (n) \rvert = A =1$ are kept constant throughout this study.

\section{Results and Analysis}

In this section, the simulation results are presented and briefly analyzed. Considering that the RISs can be configured in either of the two modes:

\begin{itemize}
    \item Fixed RIS configuration: the RIS configuration is constant whatever the MS location within the target area. Actually, the UCs are tuned such that the beam is oriented towards one specific location in the area (MS25 for RIS1, and MS36 for RIS2).
    \item MS-specific RIS configuration: each location in the target area is considered as a potential position for the MS of interest or where the user can be. Then, the RIS is optimally re-configured towards each of those positions. The result shows the maximum achievable received (RX) power for an MS moving in a specific area.
\end{itemize}

The coverage in target areas A1 and A2 is studied by setting up a grid of MSs, as shown in Fig. \ref{1.5}, with a resolution of 2 m. The total RX power is calculated by coherently summing the RIS contributions to the BS-MS paths.

\subsection{Simulation Results at 3.7 GHz}

\begin{figure}[ht!]
\centering
\includegraphics[width=2.9in]{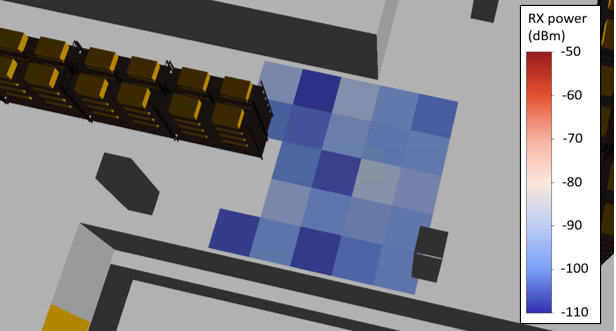}
\caption{RX power coverage map of A1 w/o RIS.}
\label{1.7}
\end{figure}

Fig. \ref{1.7} shows the RX power (dBm) coverage map of A1 at 3.7 GHz without the presence of RIS, which is the baseline situation of our study. Then, RIS1 is deployed to see the coverage enhancement it can provide in A1. 

\begin{figure}[ht!]
\centering
\includegraphics[width=3in]{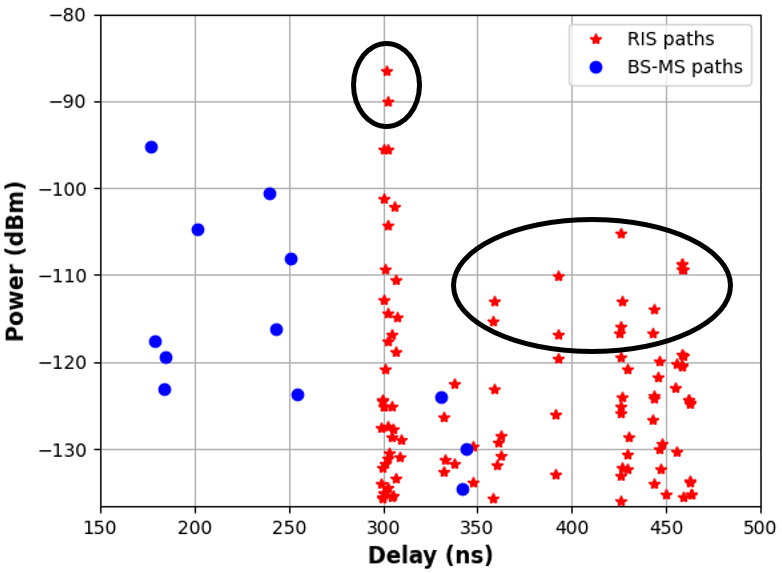}
\caption{PDP for BS1-MS25 in presence of RIS1.}
\label{1.6}
\end{figure}

Fig. \ref{1.6} shows the power delay profile (PDP) for MS25 at 3.7 GHz for 16$\times$16 RIS1. The red markers indicate the paths added just because of RIS1. We observe that the radio channel is significantly enriched, with many new long-delay echoes, and two dominant paths around 300 ns. The power of the strongest RIS-based path is around 7 dB greater than the power of the strongest conventional (BS-MS link) path.

\begin{figure}[ht!]
\centering
  \subfigure[Fixed RIS configuration.]{%
    \includegraphics[width=2.9in]{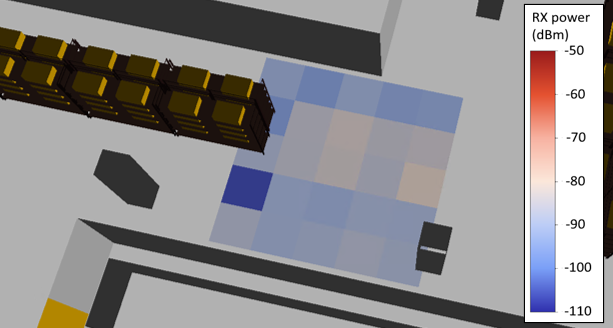}%
    \label{1.8.1}%
  } \\
  \subfigure[MS-specific RIS configuration.]{%
    \includegraphics[width=2.9in]{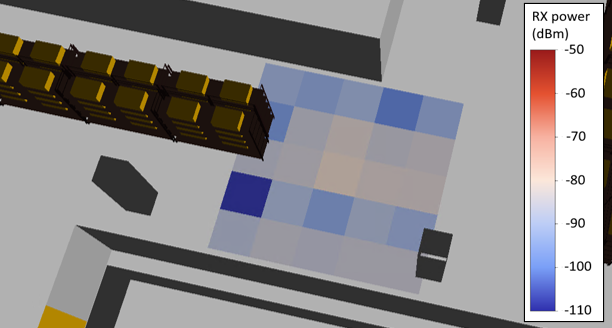}%
    \label{1.8.2}%
  }
  \caption{Total RX power coverage maps of A1 for 16$\times$16 RIS1.}
  \label{1.8}
\end{figure}

Fig. \ref{1.8.1} and Fig. \ref{1.8.2} show the total RX power coverage map of A1 for 16$\times$16 RIS1 for fixed RIS configuration and MS-specific RIS configuration, respectively. It can be observed by comparing Fig. \ref{1.7} and Fig. \ref{1.8} that 16$\times$16 RIS1 provides a coverage enhancement over the case of no RIS, with an increase of about 11 dB in the mean of the total RX power. The results with MS-specific RIS configuration show the upper bound of the total RX power in A1.

\begin{figure}[ht!]
\centering
  \subfigure[Fixed RIS configuration.]{%
    \includegraphics[width=2.9in]{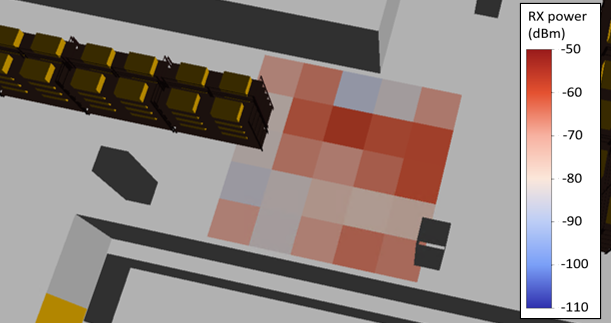}%
    \label{1.9.1}%
  } \\
  \subfigure[MS-specific RIS configuration.]{%
    \includegraphics[width=2.9in]{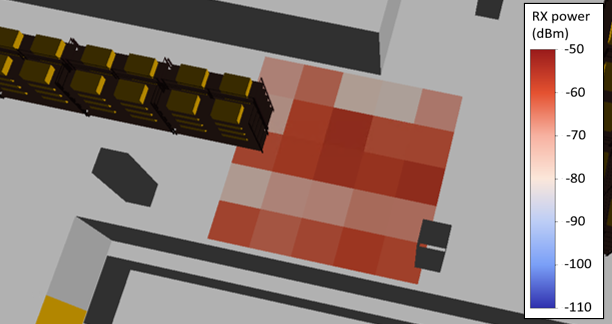}%
    \label{1.9.2}%
  } 
  \caption{Total RX power coverage maps of A1 for 32$\times$32 RIS1.}
  \label{1.9}
\end{figure}

Fig. \ref{1.9.1} and Fig. \ref{1.9.2} show the total RX power coverage maps of A1 for 32$\times$32 RIS1. Comparing Fig. \ref{1.8} and Fig. \ref{1.9} shows that a 32$\times$32 RIS provides significantly better coverage enhancement than a 16$\times$16 RIS. The mean of the total RX power increases by 17 dB and 23 dB, for fixed-RIS configuration and MS-specific RIS configuration, respectively, when the size of RIS1 is increased from 16$\times$16 to 32$\times$32. 

We were expecting the RX power difference between the 32$\times$32 RIS setup and the 16$\times$16 RIS setup to be lower or close to 12 dB since the BF gain (in linear scale) is proportional to the square of UCs number \cite{tang2022path}. This is verified when there is just a direct (LoS) link between BS-RIS and RIS-MS or when performing incoherent summation of the multiple paths (MPs); but not in the case of coherent summation of the MPs, as considered in our study. Actually, we observed at some locations that the 16$\times$16 RIS generates two or more strong propagation paths with destructive interference, which leads to almost no benefit in the total RX power (e.g., only 1 dB at MS25). This behavior is attenuated when using the 32$\times$32 RIS due to better spatial discrimination; the BF is optimally oriented for the strongest path but with lower amplification for the concurrent paths. Finally, the 32$\times$32 RIS brings both an advantage in terms of BF gain and fading reduction (e.g., 20 dB additional gain at MS25). Of course, this observation is specific to the assumptions we made: narrowband signal; RIS configuration based on the analysis of a single strongest path. The conclusions would be different for another RIS controller algorithm or for a wideband signal.

\subsection{Simulation Results at 27 GHz}

\begin{figure}[ht!]
\centering
\includegraphics[width=3.4in]{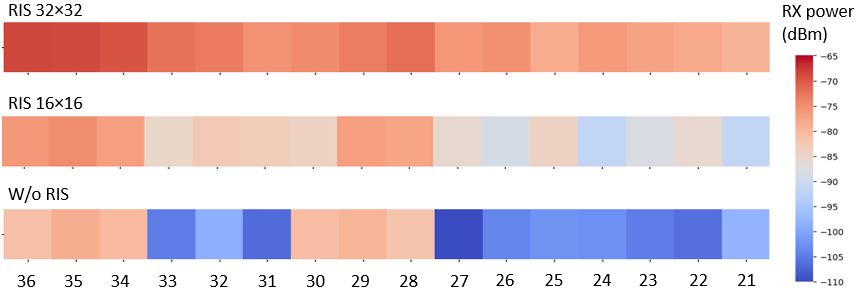}
\caption{RX power coverage maps of A2: w/o RIS (bottom), with 16$\times$16 RIS2 (middle), with 32$\times$32 RIS2 (top).}
\label{1.10}
\end{figure}

Fig. \ref{1.10}(bottom) shows the RX power coverage map of A2 without RIS. Then Fig. \ref{1.10}(middle) shows the total RX power when 16$\times$16 RIS2 is deployed. Finally, Fig. \ref{1.10}(top) shows the total RX power coverage map when 32$\times$32 RIS2 is deployed. The coverage results provided by fixed RIS configuration are almost the same for MS-specific RIS configuration since RIS2 and all MSs forming A2 are collinear, which is expected. Hence, the results shown in Fig. \ref{1.10}(middle) and Fig. \ref{1.10}(top) are for the MS-specific RIS configuration. It can be seen that 16$\times$16 RIS2 provides a coverage enhancement over the case of no RIS, with an increase of about 11 dB in the mean of the total RX power. Then 32$\times$32 RIS2 offers even better coverage enhancement than 16$\times$16 RIS2. The mean of the total RX power increases by around 10 dB when the size of RIS2 is increased from 16$\times$16 to 32$\times$32. 

\vskip-0.2\baselineskip

\subsection{CDF Analysis}

\begin{figure}[ht!]
\centering
  \subfigure[CDF of RX power for 32$\times$32 RIS1 at 3.7 GHz.]{%
    \includegraphics[width=3.4in]{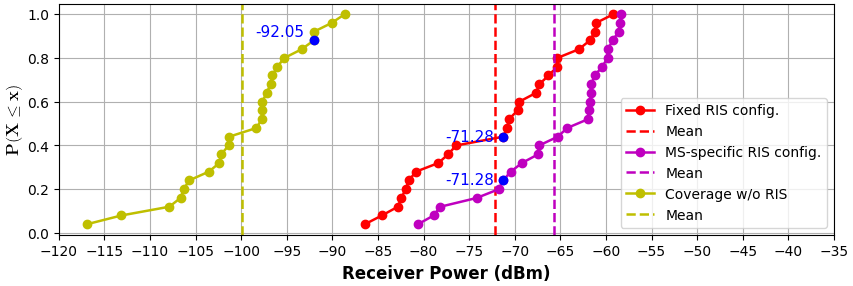}%
    \label{1.11.1}%
  } \\
  \subfigure[CDF of RX power for 32$\times$32 RIS2 at 27 GHz.]{%
    \includegraphics[width=3.4in]{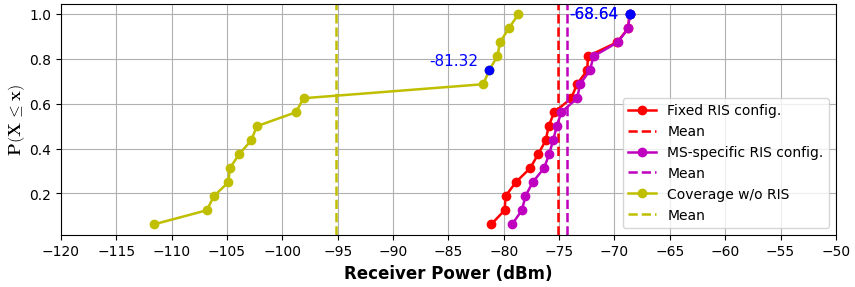}%
    \label{1.11.2}%
  }
  \caption{CDF of total RX power at 3.7 GHz and 27 GHz.}
  \label{1.11}
\end{figure}

Fig. \ref{1.11.1} and Fig. \ref{1.11.2} show the CDF of RX powers obtained with the 32$\times$32 RIS for 3.7 GHz and 27 GHz scenarios, respectively. Table \ref{tab2} compares different performance parameters extracted from the CDFs shown in Fig. \ref{1.11}. In the absence of any RIS, the minimum RX power found in the RX area is the consequence of a severe obstruction loss and possibly some destructive multi-path summation; this minimum RX power is increased by at least 30 dB when adding the RIS, which is of critical importance to get seamless coverage.

Table \ref{tab2} also shows the mean gains (increase in the mean RX power when using the RIS as compared to no RIS usage) offered by the RIS within the target areas, for both studied RIS configurations. This gain depends on many parameters, such as the distances between BS-RIS and RIS-MS; but it is worth noting that a 32$\times$32 RIS can bring more than 20 dB (or possibly more than 30 dB) average gain in a realistic factory NLoS FF situation. The coverage rate shown in Table \ref{tab2} is calculated as the percentage of locations where RX power is greater than a given threshold. This threshold is usually determined as a function of the receiver performance and target user service. Here, for simplicity, we considered a threshold of -105 dBm for access to minimum service and -80 dBm for high throughput applications. It can be seen that using an RIS can bring more than 70\% coverage rate (or even 100\% in some cases) in the shadowed areas.

\vskip-0.4\baselineskip
\begin{table}[ht!]
    \centering
    \caption{Comparison of Coverage Rate and Mean Gain}
    \label{tab2}
    \begin{tabular}{|>{\centering\arraybackslash\vspace{-0.3ex}}m{2.6cm}<{\vspace{0ex}}|>{\centering\arraybackslash\vspace{-0.3ex}}m{1.5cm}<{\vspace{0ex}}|>{\centering\arraybackslash\vspace{-0.3ex}}m{1.5cm}<{\vspace{0ex}}|>{\centering\arraybackslash\vspace{-0.3ex}}m{1.4cm}<{\vspace{0ex}}|}
        \hline
        \textbf{RIS configuration (Frequency)} & \textbf{Coverage Rate (\%) @-80 dBm} & \textbf{Coverage Rate (\%) @-105 dBm} & \textbf{Mean Gain (dB)}\\
        \hline
        no RIS (3.7 GHz) & 0\% & 76\% & - \\ 
        \hline
        Fixed (3.7 GHz) & 72\% & 100\% & 27.8 \\ 
        \hline
        MS-specific (3.7 GHz) & 96\% & 100\% & 34.1 \\ 
        \hline
        no RIS (27 GHz) & 8\% & 88\% & - \\ 
        \hline
        Fixed (27 GHz) & 96\% & 100\% & 20.1 \\ 
        \hline
        MS-specific (27 GHz) & 100\% & 100\% & 20.9 \\ 
        \hline
    \end{tabular}
\end{table}

\section{Conclusion and future perspective}

In this paper, we integrated the RIS scattering model in an RT simulator. This model relies on an accurate UC scattering function that was selected after a comparison of different options available in the literature; completed with a simple automated RIS reconfiguration method. The resulting ray-based system-level tool adds several RIS-based propagation paths to the conventional multi-paths. It was then used to demonstrate RIS-aided coverage enhancement in a highly cluttered warehouse of a factory. The coverage maps, CDF of total RX powers, and performance parameters such as coverage rate and mean gain were analyzed in order to investigate and quantify RIS-aided coverage extension in the factory scenario at 3.7 GHz and 27 GHz. It was demonstrated that deploying an RIS can provide strong coverage enhancements in some shadowed areas in a factory setting.

Specific target areas and RIS sizes were chosen for this study to demonstrate RIS-aided coverage enhancement; however, shorter distances, larger areas, and bigger RISs could not be considered in order to satisfy the FF conditions and reasonable RT computation timing. This work is already being extended to RIS near-field and improve RT computation timing \cite{EuCNC1} to demonstrate RIS-aided coverage in the entire area. We expect that the preliminary results presented in this paper will be consolidated, which show significant coverage enhancements by deploying RIS in a factory scenario.



\section*{Acknowledgment}

This work is part of a project that has received funding from the European Union's Horizon 2020 research and innovation programme under the Marie Skłodowska Curie grant agreement No. 956670.


\bibliographystyle{IEEEtran}
\bibliography{IEEEabrv, mainbib}

\end{document}